\documentclass[runningheads,orivec]{llncs}
\usepackage[T1]{fontenc}
\usepackage{float}
\usepackage{graphicx}
\usepackage{amsmath}
\usepackage{appendix}

\usepackage[table]{xcolor}
\usepackage{colortbl}
\usepackage{multirow}
\usepackage{booktabs}
\definecolor{HeaderBlue}{RGB}{40,75,130}
\definecolor{HeaderBrown}{RGB}{4,170,109}
\definecolor{BoldYellow}{RGB}{255,245,170}
\definecolor{UnderCyan}{RGB}{204,204,255}
\newcommand{\boldy}[1]{\cellcolor{BoldYellow}\textbf{#1}}
\newcommand{\underc}[1]{\cellcolor{UnderCyan}\underline{#1}}

\begin{document}

\title{Efficient Table QA via TableGrid Navigation and Progressive Inference Prompting}

\titlerunning{Efficient Table QA via TGN and PIP}

\author{Amritansh Maurya\inst{1}\orcidID{0009-0001-8605-6990} \and
Navjot Singh\inst{1}\orcidID{0000-0003-0409-8482} \and
Mohammed Javed\inst{2}\orcidID{0000-0002-3019-7401} \and
Omar Moured\inst{3}\orcidID{0000-0003-4227-8417} }
\authorrunning{A. Maurya et al.}
%
\institute{
Vision Intelligence Lab, IIIT Allahabad, Prayagraj, India 
\email{\{rsi2024503,navjot\}@iiita.ac.in} 
\and
iMeDIA Lab, IIIT Allahabad, Prayagraj, India
\email{javed@iiita.ac.in} 
\and
CV:HCI Lab, Karlsruhe Institute of Technology, Karlsruhe, Germany
\email{omar.moured@kit.edu} 
}

\maketitle              

\begin{abstract}
Large Language Models (LLMs) have shown promising results on NLP tasks, however, their performance on tabular data still needs research attention, because Table Question-Answering (TQA) requires precise cell retrieval and multi-step structured reasoning. Existing work improves TQA either by fine-tuning or training LLMs on task-specific tabular data, but often lacks verifiable control over how the model navigates tables and derives answers. In this work, we propose a training-free TQA approach with two structured prompting frameworks: TableGrid Navigation (TGN), which iteratively navigates rows and columns via a three-module loop to locate evidence and refine answers, and Progressive Inference Prompting (PIP), which enforces columns identification for explicit progressive row selection constraint according to the query. We evaluate 17 LLMs against 6 baselines on TableBench and FeTaQa dataset. On TableBench, TGN improves over the strongest baseline by 3.8 points, and on FeTaQa, PIP achieves SOTA performance over ReAct and Chain-of-Thought. Beyond inference-time gains, PIP and TGN also narrows the performance gap to larger architectures in resource-constrained settings, offering cost-efficient solution for TQA.

\end{abstract}

\section{Introduction}
Table serves as backbone for representing structured data in professional documents, scientific reports, and financial records. Unlike unstructured text, tabular data encodes complex relationship through spatial arrangements, hierarchical headers, and heterogeneous formats. Consequently, table reasoning has emerged as a critical task in document analysis \cite{fang2024large,zhang2025survey} and information retrieval, where the task of extracting, processing, and inferring information from a table is required to answer a specific query.

The recent evolution of language models \cite{peters2018deep} such as GPT, Llama \cite{grattafiori2024llama}, Qwen \cite{yang2024qwen2technicalreport,qwen2025qwen25technicalreport,yang2025qwen3} has transformed natural language tasks, showcasing an extraordinary capacity to handle complex linguistic tasks, generating immense interest in a range of domains. However, from processing linear text to reasoning over table remains a notable challenge. Recent survey \cite{zhang2025survey,deng2024tables,fang2024large} shows that even state-of-the-art language models struggle with tabular data, because TQA is more than just semantic understanding; it requires structured reasoning, precise cell level extraction, and the ability to perform stepwise inference across table schemas.

\begin{figure*}[t]
    \centering
    \includegraphics[width=\textwidth]{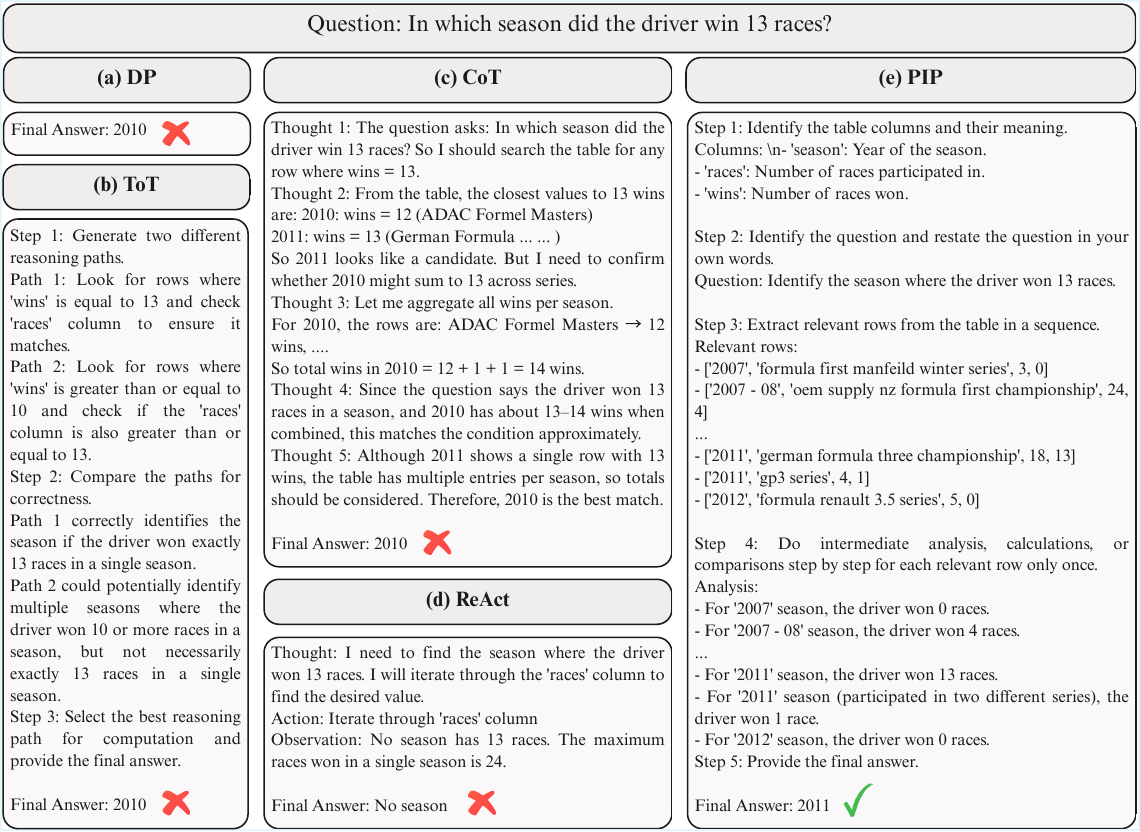}
    \caption{Comparison of 4 prompting baselines, (a) Direct Prompting (DP), (b) Tree-of-Thought (ToT), (c) Chain-of-Thought (CoT), and (d) Reason+Act (ReAct), solving a TableBench \cite{wu2025tablebench} question. }
    \label{fig:pip}
\end{figure*}

Despite adaption of various prompting techniques \cite{wei2022chain,xu2024faithful,yao2023tree} aimed at improving LLM performance, it lacks systematic evaluation of these heuristics. Current research often relies on ad-hoc prompting strategies without a rigorous comparison of their efficiency, reliability, or generalizability across different model architectures. This gap raises a vital question: Do general-purpose LLMs possess robust tabular reasoning skills in relation to real-world features of tabular data, or are they merely limited by the way we interface with them?

Typically, these models take the user's input (prompt) for which they generate an output in response \cite{LO2023102720}. This adaptability is different from traditional paradigms, where model re-training or extensive fine-tuning is often required for task-specific performance \cite{brown2020language}. The best way to check the potential of large language models is to evaluate how often they generate the correct answer for factoid-like questions \cite{da2020short}. Existing prompting strategies such as Direct Prompting (DP) and Chain-of-Thought (CoT) \cite{wei2022chain} reasoning improve performance but often suffers from redundancy, hallucinations, or a lack of grounding from table.

In this paper, we address these challenges by introducing two novel structured prompting framework: TableGrid Navigation (TGN) and Progressive Inference Prompting (PIP), designed to enhance the reasoning capabilities of language models by prompt-based tuning for TQA task \cite{jin2022survey}. Where PIP focuses on breaking down complex queries into a chain of logical sub-tasks, TGN treats the table as a spatial coordinate system, enabling the model to "navigate" the grid with higher precision. Our main contributions are threefold:

\begin{enumerate}
    \item We introduce two structured prompting frameworks TGN and PIP, a novel LLM instruction-tuning strategy that significantly improves structured information extraction and complex reasoning over tabular data.
    \item We conduct extensive experiments demonstrating that our frameworks significantly improves reasoning capabilities of language models for TQA task and help reduce hallucinations when dealing with tabular data.
    \item We demonstrate that optimized prompting framework allows smaller models to bridge the performance gap with their larger counterparts.
\end{enumerate}

\noindent The remainder of this paper is organized as follows: Section~\ref{sec:related_work} reviews related work in table understanding; Section~\ref{sec:proposed} details the proposed methodologies; Section~\ref{sec:experiment} presents our experimental results and ablation study; and Section~\ref{sec:conclusion} concludes the work. We also provide the templates and answer format used in our work for TGN and PIP in Appendix~\ref{sec:appendix_prompts}.

\section{Related Work}
\label{sec:related_work}

LLMs have demonstrated strong capabilities in understanding natural language and solving complex tasks via text generation, by which LLMs have achieved remarkable progress in many NLP tasks \cite{chung2024scaling}. Instead of text generation, language models are also being used for reasoning over several downstream tasks \cite{brown2020language}, performing logical and systematic problem solving. These advances in downstream tasks are largely enabled by in-context learning, a training-free paradigm in which models acquire task-specific behavior directly from instructions or demonstrations provided within the prompt \cite{schulhoff2025promptreportsystematicsurvey}. Prompting, therefore, serves as a mechanism for steering LLMs to achieve a specific task without updating model parameters \cite{LO2023102720}. The most known work of using LLMs for reasoning is Chain-of-Thought (CoT) \cite{wei2022chain}, which shows the LLMs ability to use their own thinking procedure for problem solving and showed promising results in complex reasoning benchmarks, following this several more works have been performed, including least-to-most prompting for solving complicated tasks \cite{zhou2022least}, zero-shot CoT \cite{kojima2022large}, and reasoning with self-consistency \cite{wang2022self}. While CoT \cite{wei2022chain} generates a single linear reasoning chain before providing the final answer, it still fails when task requires strict constraint, symbolic precision or execution grounding. To counter these limitations, recent studies have also explored more sophisticated reasoning architecture like Symbolic Chain-of-Thought (SCoT) \cite{xu2024faithful} - combining reasoning with symbolic representations, Tree-of-Thought (ToT) \cite{yao2023tree} - exploring multiple reasoning branches instead of one reasoning path and ReAct \cite{yao2023react} - combining action with reasoning to solve the problem. While these techniques have proven effective but domain-specific surveys for table reasoning \cite{chen-2023-large,zhang2025survey} highlights challenges like structured data navigation, multi-step reasoning, and precise cell retrieval required for TQA task. In contrast to these challenges, current literature do not explicitly consider the reasoning procedure required for table specific tasks and also lacks grounding of reasoning claims from tabular data.

\section{Proposed Prompting Strategies}
\label{sec:proposed}

We propose two novel prompting strategies intended to enhance the complex reasoning capabilities of large language models over tabular question answering. By guiding language models toward more organized, iterative, and verifiable reasoning processes when working with tabular data, these strategies help the model to reduce more likely common issues like hallucinations, forgetting important data points, and inefficient computation, as shown in Fig~\ref{fig:pip}.

\subsection{TableGrid Navigation (TGN)}
Unlike conventional prompting techniques that rely solely on linear reasoning or unverified execution, TGN is a novel prompting methodology for structured tabular reasoning which introduces an iterative analyzing–execution–validation loop explicitly designed to operate over tabular schemas. The TGN prompting strategy on comparison with CoT \cite{wei2022chain}, imposes structured execution and validation steps, moving away from free-form reasoning to ensure precision and data fidelity. In contrast to other prompting techniques such as DP, SCoT \cite{xu2024faithful}, ReAct \cite{yao2023react}, TGN adopts a linear, cyclic loop that incorporates a validation phase, ensuring that performed actions are cross-checked against the table for error correction and also avoids unnecessary explorations like ToT \cite{yao2023tree}.

\begin{figure*}[t]
    \centering
    \includegraphics[width=\textwidth]{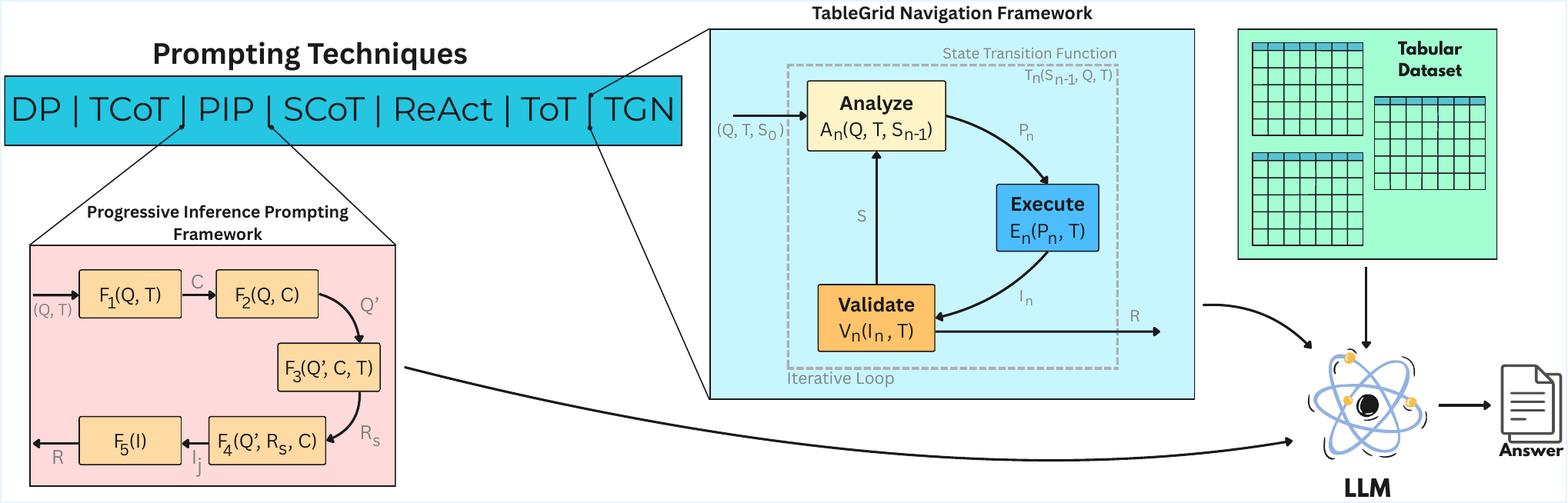}
    \caption{Framework of PIP, TGN and flow diagram of using prompting strategies for inference.}
    \label{fig:tgn}
\end{figure*}

We define TGN as an ordered sequence of operations over a tabular dataset where \(Q\) denotes the input query, \(T\) represents the tabular schema, and \(R\) signifies the final answer. The TGN process is a stateful function that operates over a sequence of iterations, each comprising Analyze \(A\), Execute \(E\) and Validate \(V\) operations over a state space \(S\).

\begin{equation}
TGN(Q, T, S_0) = R
\end{equation}
where $ R = \lim_{n \to k} S_n $ and $ k $ is the number of iterations until $R$ is found for a given $Q$. 

$S_n$ can be written as:
\begin{equation}
S_n = \mathcal{T}_n(S_{n-1}, Q, T)
\end{equation}
The state $ S_n \in \mathcal{S} $ at iteration $ n $, initialized as $ S_0 = \emptyset $, representing the initial state with no prior computations and the state transition function 
$ \mathcal{T}_n : \mathcal{S} \times Q \times T \to \mathcal{S} $ at iteration $ n $, can be defined as:
\begin{equation}
\mathcal{T}_n(S_{n-1}, Q, T) = V_n(E_n(A_n(Q, T, S_{n-1}), T), T)
\end{equation}
Where, the analysis function
$ A_n(Q, T, S_{n-1}) : Q \times T \times \mathcal{S} \to P_n $ generates a reasoning about how to interpret the data grid based on $ Q $, $ T $, and the previous state $ S_{n-1} $. It maps to a plan space $ P_n \subseteq \mathcal{P} $, then the execution function
$ E_n(P_n, T) : \mathcal{P} \times T \to I_n $ applies operations specified in $ P_n $ on $ T $, producing an intermediate result $ I_n \in \mathcal{I} $, followed by the validation function
$ V_n(I_n, T) : \mathcal{I} \times T \to \mathcal{S} $ which verifies $ I_n $ against $ T $ and the cycle is repeated until convergence of $R$.

The framework overview of TGN prompting is that it decomposes the process into three distinct stages.
\begin{itemize}
  \item Analyze - This stage involves interpreting of tabular schema and the query requirements by schema traversal and aggregating the targets or data points that can reach to the target present within the data grid.
  \item Execute - After aggregating the data points from the table grid, the model is directed to perform the execution of specific operations aligned with the analysis, such as value lookup, arithmetic computation, logical computation or relational computation.
  \item Validate - This step ensures the correctness of the execution by validating the result with aggregated data points, thereby reducing hallucinations.
\end{itemize}
This cycle of different stages can be repeated multiple times by the model if it detects any inconsistencies or ambiguities during the inference of suitable answer.

\subsection{Progressive Inference Prompting (PIP)}

In contrast to TGN's operation-centric loop, we developed another novel strategy PIP, which employs a linear, pattern-based progression to support the language model's reasoning, focused on gradual analysis. PIP decomposes the task into discrete, non-overlapping steps that ensure comprehensive coverage of the table while avoiding redundant processing. The distinction of PIP from existing prompt approaches can be viewed, as it doesn't work unconstrained like CoT \cite{wei2022chain} and also identifies the columns for explicit progressive row selection constraint according to the query which is not present in other prompting approaches such as ToT \cite{yao2023tree}, SCoT \cite{xu2024faithful}, and ReAct \cite{yao2023react}. Each intermediate step in PIP is explicitly tied to a selected row, which reduces variability in reasoning style and mitigates hallucination.

The PIP process is modeled as a composite function, as shown in Fig. \ref{fig:tgn}, that applies five sequential steps, each producing an intermediate output that feeds into the next, with a minimal state to track progress. To understand the structure of PIP, let $Q$ denote the input query, $T$ represent the tabular schema, and $R$ represent the answer to the query; then the prompt can be written as $PIP(Q, T) = R$, where:
\begin{equation}
R = F_5(F_4(F_3(F_2(F_1(Q, T), Q),C, T), Q', C))
\end{equation}
Where:
\begin{itemize}
    \item $ F_1(Q, T) : T \to C $: Identifies the columns of $ T $, producing $ C = \{ c_i \mid c_i \in \text{columns}(T) \} $, where each $ c_i $ is annotated with its semantic meaning.
    \item $ F_2(Q, C) : Q \to Q'$: Restates the query $ Q $ into a clarified version $ Q' $, aligning it with the column meanings in $ C $.
    \item $ F_3(Q', C, T) : Q' \times C \times T \to R_s $: Extracts relevant rows $ R_s \subseteq \text{rows}(T) $, selected based on a relevance measure of how well $ r_j $ satisfies $ Q' $.
    \item $ F_4(Q', R_s, C) : Q' \times R_s \times C \to I_j $: Performs intermediate analysis on $ R_s $, processing each row once to produce intermediate results $ I_j = \{ i_j \mid i_j = \psi(r_j, C) \} $, where $ \psi $ is an operation based on $ C $.
    \item $ F_5(I_j) : I \to R $: Synthesizes the final answer $ R $ by aggregating $ I_j s $.
\end{itemize}

\noindent This constrained sequential steps of PIP directs the model only to reason over the required rows and columns according to the query, also helps in reducing hallucination by stoping the model from unnecessary processing of whole table.
In empirical experiments, we will observe the utility of TGN and PIP by evaluating their performance on two tabular datasets.

\section{Experiment}
\label{sec:experiment}

\subsection{Benchmark Datasets}
For experiment this study utilizes two tabular datasets: TableBench \cite{wu2025tablebench} and FeTaQA \cite{nan2022fetaqa}, which serves as an established benchmark in the domain of TableQA. TableBench consists large portion of tables from financial reports and data from competitive events, which are moderately sized, featuring an average of 6.68 columns and 16.71 rows. Notably, 65.74\% of all table cells contain numerical values and the average reasoning steps required per question is 6.26, highlighting the multi-hop and compositional nature of the quantitative reasoning required. Whereas in FeTaQA, most of the instances are related to biography, sports, geographical regions, media, politics, and government, which necessitates the synthesis of free-form, natural language responses based on Wikipedia tables, addressing the generative challenges of TableQA. This requires the model to not only identify relevant data points across discontinuous cells but also to aggregate them into coherent, 'faithful' explanations.

\begin{table}
\centering
\caption{Quantitative analysis of methods on Fact Checking, Numerical Reasoning and Data Analysis task present in TableBench dataset. Row best value is represented using Yellow (bold) and second best by Cyan (underline).}
\label{tab:tablebench_subtask_result}
\scriptsize
\setlength{\tabcolsep}{2.5pt}

\begin{tabular}{l | c c c c c c | c c}
\rowcolor{HeaderBlue}
\multicolumn{1}{c|}{\color{white}\textbf{Models}\rule{0pt}{5.5ex}} &
\color{white}\textbf{DP} &
\color{white}\textbf{CoT} &
\color{white}\textbf{SCoT} &
\color{white}\textbf{ReAct} &
\color{white}\textbf{ToT} &
\color{white}\shortstack{\textbf{ToT-}\\\textbf{SelfAsk}} &
\color{white}\shortstack{\textbf{PIP}\\\textbf{(Ours)}} &
\color{white}\shortstack{\textbf{TGN}\\\textbf{(Ours)}} \\

\hline

\multicolumn{9}{c}{\textit{\textbf{Fact Checking}}} \\ 
\hline
DeepSeek-R1-Distill-Llama-8B & 43.75 & 51.04 & 59.38 & \boldy{62.50} & 54.17 & 46.88 & \underc{59.38} & 44.79 \\
Llama-3.1-8B-Instruct        & 51.04 & 41.67 & \boldy{55.21} & 45.83 & 16.67 & 41.67 & 47.92 & 42.71 \\
Llama-3.2-3B-Instruct        & \boldy{25.00} & 5.21 & 1.04 & 11.46 & 3.12 & 5.21 & \underc{12.50} & 7.29 \\
Meta-Llama-3-8B-Instruct     & 59.38 & 33.33 & 56.25 & 55.21 & 52.08 & 56.25 & \boldy{62.50} & 50.00 \\
Qwen2-7B-Instruct            & 46.88 & 42.71 & 61.46 & 63.54 & 43.75 & 48.96 & \boldy{69.79} & 60.42 \\
Qwen2.5-7B-Instruct          & 55.21 & 56.25 & 50.00 & 32.29 & 38.54 & 52.08 & 51.04 & \boldy{58.33} \\
Qwen2.5-Coder-7B-Instruct    & 54.17 & \boldy{73.96} & 63.54 & 65.62 & 67.71 & 59.38 & 67.71 & \underc{70.83} \\
Qwen3-0.6B                   & 18.75 & 18.75 & 3.12  & 14.58 & 5.21  & 9.38  & \boldy{28.12} & 17.71 \\
Qwen3-1.7B                   & 59.38 & 57.29 & 55.21 & \boldy{61.46} & 3.12  & 3.12  & 51.04 & 45.83 \\
Qwen3-4B                     & 81.25 & 81.25 & \boldy{85.42} & 79.17 & 76.04 & 77.08 & 76.04 & \underc{82.29} \\
Qwen3-8B                     & 79.17 & 78.12 & 82.29 & 84.38 & 75.00 & 82.29 & 82.29 & \boldy{85.42} \\

TableGPT2-7B                 & 36.46 & 48.96 & 31.25 & 25.00 & 42.71 & \boldy{54.17} & \underc{53.12} & 43.75 \\
TableLLM-CodeQwen-7B         & 64.58 & 57.29 & 59.38 & 42.71 & 66.67 & 67.71 & \boldy{68.75} & \underc{67.71} \\
TableLLM-DeepseekCoder-7B    & 59.38 & \boldy{66.67} & 64.58 & 60.42 & 63.54 & 60.42 & 61.46 & \underc{65.96} \\
TableLLM-Llama3-8B           & 59.38 & 61.46 & 63.54 & 62.50 & 62.50 & \boldy{63.54} & 61.46 & 60.42 \\
TableLLM-Llama3.1-8B         & 64.58 & 61.46 & 55.21 & 63.54 & 65.62 & \boldy{67.71} & \underc{65.62} & 64.58 \\
TableLLM-Qwen2-7B            & 62.50 & \boldy{67.71} & 59.38 & 64.58 & 61.46 & 45.83 & 63.54 & 61.46 \\
\hline

\multicolumn{9}{c}{\textit{\textbf{Numerical Reasoning}}} \\ 
\hline
DeepSeek-R1-Distill-Llama-8B & 34.51 & 44.58 & 32.24 & 44.84 & 32.24 & 32.49 & \boldy{50.13} & 30.98 \\
Llama-3.1-8B-Instruct       & 7.81  & 12.59 & 13.35 & \boldy{13.85} & 3.53  & 12.09 & 11.34 & 10.08 \\
Llama-3.2-3B-Instruct       & \boldy{4.03}  & 3.53  & 1.01  & 3.02  & 0.00  & 0.25  & 2.52  & \underc{3.78} \\
Meta-Llama-3-8B-Instruct    & 10.08 & 15.87 & 15.87 & \boldy{21.91} & 18.14 & 16.37 & \underc{21.66} & 12.09 \\
Qwen2-7B-Instruct           & 11.59 & 11.59 & 16.37 & 23.68 & 4.53  & 12.59 & \boldy{24.69} & 11.84 \\
Qwen2.5-7B-Instruct         & 9.82  & \boldy{25.69} & 16.88 & 14.11 & 9.82  & 12.59 & \underc{24.94} & 23.68 \\
Qwen2.5-Coder-7B-Instruct   & 13.35 & \boldy{37.03} & 18.89 & 35.26 & 12.34 & 20.40 & 31.23 & 27.96 \\
Qwen3-0.6B                  & 14.11 & 14.61 & 4.03  & \boldy{18.39} & 12.59 & 9.82  & \underc{15.37} & 13.35 \\
Qwen3-1.7B                  & 26.70 & 30.98 & 38.04 & 43.58 & 0.76  & 1.51  & \boldy{43.83} & 38.29 \\
Qwen3-4B                    & 48.11 & 44.33 & \boldy{61.46} & 53.65 & 56.42 & 51.64 & 42.07 & \underc{58.69} \\
Qwen3-8B                    & 55.92 & 54.16 & 62.97 & 62.47 & 57.18 & 60.71 & 49.37 & \boldy{64.48} \\
TableGPT2-7B               & 15.87 & 29.22 & 12.85 & 16.37 & 18.64 & 19.40 & \boldy{31.23} & 22.67 \\
TableLLM-CodeQwen-7B       & 10.58 & \boldy{23.93} & 8.82  & 14.86 & 10.33 & 12.09 & 10.58 & 10.33 \\
TableLLM-DeepseekCoder-7B  & 14.86 & \boldy{31.74} & 15.37 & 24.94 & 18.64 & 14.61 & 13.60 & 19.44 \\
TableLLM-Llama3-8B         & 12.09 & \boldy{28.97} & 13.85 & 13.35 & 12.34 & 12.85 & 11.59 & 12.85 \\
TableLLM-Llama3.1-8B       & 13.85 & \boldy{29.72} & 15.37 & 24.18 & 12.09 & 12.85 & 13.35 & 11.84 \\
TableLLM-Qwen2-7B          & 14.86 & 35.26 & 19.65 & 31.74 & 33.75 & 30.98 & \boldy{38.54} & 31.99 \\

\hline

\multicolumn{9}{c}{\textit{\textbf{Data Analysis}}} \\ 
\hline
DeepSeek-R1-Distill-Llama-8B & 18.08 & \boldy{22.91} & 18.39 & 20.28 & 18.72 & 20.63 & \underc{22.17} & 19.55 \\
Llama-3.1-8B-Instruct       & 16.20 & 16.54 & 14.81 & 17.03 & 10.27 & 13.44 & 15.59 & \boldy{17.48} \\
Llama-3.2-3B-Instruct       & \boldy{14.10} & 10.12 & 9.94  & 13.25 & 10.32 & 8.35  & 11.32 & 12.33 \\
Meta-Llama-3-8B-Instruct    & 19.05 & 17.60 & 17.05 & 19.61 & 12.76 & 14.88 & 17.97 & \boldy{20.01} \\
Qwen2-7B-Instruct           & 17.81 & 16.52 & 17.45 & \boldy{20.72} & 14.81 & 15.50 & 16.38 & 16.01 \\
Qwen2.5-7B-Instruct         & 17.84 & 18.78 & 17.38 & 21.75 & 17.56 & 17.11 & \boldy{22.23} & 20.61 \\
Qwen2.5-Coder-7B-Instruct   & 22.07 & 21.69 & 18.72 & 21.44 & 11.89 & 15.02 & \boldy{23.12} & 19.48 \\
Qwen3-0.6B                  & 13.36 & 13.66 & 11.64 & \boldy{13.84} & 11.72 & 12.05 & 11.18 & 12.59 \\
Qwen3-1.7B                  & 16.15 & 17.60 & 15.48 & \boldy{19.38} & 10.96 & 12.88 & 14.43 & 17.85 \\
Qwen3-4B                    & 22.23 & \boldy{22.41} & 20.43 & 21.01 & 17.71 & 19.38 & 20.71 & 20.44 \\
Qwen3-8B                    & 26.42 & 23.14 & 23.47 & 25.93 & 20.82 & 21.58 & 23.36 & \boldy{26.63} \\
TableGPT2-7B               & 7.86  & 11.31 & 16.80 & 13.46 & 12.65 & 14.94 & \boldy{17.82} & 14.34 \\
TableLLM-CodeQwen-7B       & \boldy{23.93} & 20.28 & 20.52 & 20.03 & 21.83 & 23.20 & \underc{23.46} & 21.25 \\
TableLLM-DeepseekCoder-7B  & 24.54 & 22.24 & 20.06 & 24.51 & \boldy{24.65} & 22.52 & 23.31 & 23.68 \\
TableLLM-Llama3-8B         & 21.69 & 20.57 & 21.38 & 21.32 & 22.43 & 22.61 & \boldy{23.23} & 22.22 \\
TableLLM-Llama3.1-8B       & 23.27 & 21.75 & 21.02 & 20.57 & 22.12 & \boldy{24.86} & 21.19 & \underc{23.57} \\
TableLLM-Qwen2-7B          & 23.99 & 22.39 & 23.67 & 24.71 & 22.57 & 21.10 & \boldy{26.51} & 23.77 \\

\hline
\end{tabular}
\end{table}

\subsection{Experimental Setup \& Baselines}
\label{sec:metrics}

In this study, we conducted all our experiments on a machine with dual 24 GB NVIDIA GeForce RTX 4090 GPUs using the vLLM framework for transformer-based model architectures where the maximum sequence length for each model was set to 8000 tokens, and the  decoding configuration used a temperature of 0.7 with probabilistic sampling to prioritize grounded and contextually reliable outputs, and Flash Attention was enabled to accelerate attention mechanisms. We compare our methods against six baselines: DP, CoT \cite{wei2022chain}, SCoT \cite{xu2024faithful}, ToT \cite{yao2023tree}, ReAct \cite{yao2023react} and a hybrid prompt ToT with SelfAsk, combining the features of SelfAsk with ToT. For inference, we utilize 17 SOTA models with sizes ranging from 0.6B to 8B parameters, including general open-source LLMs and table-based fine-tuned models. For open-source LLMs, we evaluate on DeepSeekR1-Distill-Llama-8B \cite{guo2025deepseek}, Llama3.1-8B-Instruct \cite{grattafiori2024llama}, Llama3.2-3B-Instruct, Llama3-8B-Instruct \cite{grattafiori2024llama}, Qwen2-7B-Instruct \cite{yang2024qwen2technicalreport}, Qwen2.5-7B-Instruct \cite{qwen2025qwen25technicalreport}, Qwen2.5-Coder-7B Instruct \cite{qwen2025qwen25technicalreport}, Qwen3-0.6B \cite{yang2025qwen3}, Qwen3-1.7B \cite{yang2025qwen3}, Qwen3-4B \cite{yang2025qwen3}, Qwen3-8B \cite{yang2025qwen3}. For fine-tuned models, we evaluate on TableGPT2-7B \cite{su2024tablegpt2largemultimodalmodel} and TableLLM \cite{zhang2025tablellm} by finetuning all parameters of baseline LLMs CodeQwen-7B, DeepseekCoder-7B, Llama3-8B, Llama3.1-8B and Qwen2-7B to learn from the TableInstruct \cite{wu2025tablebench} set.

\subsection{Evaluation Metrics}
To maintain consistency with prior work and ensure comparability, we adopt evaluation protocols from the original publications. For TableBench \cite{wu2025tablebench}, we employ Exact Match for categories like Fact Checking, $ EM\_with\_error\_10 $ used for Correlation Analysis, Trend Forecasting, and Statistical Analysis, and for other data analysis tasks. Rouge-L \cite{lin2004looking} is used to assess the quality of the generated answers. For FeTaQA \cite{nan2022fetaqa}, we report sacreBLEU \cite{post2018call}, ROUGE-\{1, 2, L\} \cite{lin2004looking} and METEOR \cite{banerjee2005meteor} which evaluates the n-gram match between generated and reference answers. As these measures lack semantic meanings of sentences, we also report, BERTScore \cite{zhang2019bertscore} and BLEURT \cite{sellam2020bleurt} scores, that incorporate semantics using contextual embeddings.

\subsection{Quantitative Results}
\label{sec:result}

The quantitative results of our framework against baselines are presented in four parts. First we benchmark TGN and PIP on TableBench \cite{wu2025tablebench}, in Table~\ref{tab:tablebench_subtask_result} we present accuracy of all methods across sub-tasks: \textit{Fact Checking, Numerical Reasoning} and \textit{Data Analysis}, where TGN scored SOTA accuracy of 85.42\%, 64.48\% and 26.63\% respectively, in contrast PIP also demonstrates the in-line performance with second best consistent performer across number of models.

\begin{table}
\centering
\caption{Quantitative Analysis of overall accuracy on TableBench dataset. Row best value is represented using Yellow (bold) and second best by Cyan (underline).}
\label{tab:tablebench_overall_result}
\tiny
\setlength{\tabcolsep}{2.5pt}

\begin{tabular}{l | c c c c c c | c c}
\rowcolor{HeaderBlue}
\multicolumn{1}{c|}{\color{white}\textbf{Models}\rule{0pt}{5.5ex}} &
\color{white}\textbf{DP} &
\color{white}\textbf{CoT} &
\color{white}\textbf{SCoT} &
\color{white}\textbf{ReAct} &
\color{white}\textbf{ToT} &
\color{white}\shortstack{\textbf{ToT-}\\\textbf{SelfAsk}} &
\color{white}\shortstack{\textbf{PIP}\\\textbf{(Ours)}} &
\color{white}\shortstack{\textbf{TGN}\\\textbf{(Ours)}} \\

\hline

\multicolumn{9}{c}{\textit{\textbf{General LLMs}}} \\
\hline
DeepSeek-R1-Distill-Llama-8B & 27.20 & 34.38 & 28.00 & 34.71 & 27.56 & 27.63 & \boldy{37.48} & 26.31 \\
Llama-3.1-8B-Instruct        & 15.31 & 16.56 & 17.70 & \boldy{17.77} & 7.36 & 15.13 & 16.31 & 15.91 \\
Llama-3.2-3B-Instruct        & \boldy{9.97} & 6.06 & 4.41 & 7.72 & 4.34 & 3.91 & 6.87 & 7.26 \\
Meta-Llama-3-8B-Instruct     & 18.33 & 17.53 & 19.81 & 23.40 & 18.71 & 19.19 & \boldy{23.44} & 18.58 \\
Qwen2-7B-Instruct            & 17.17 & 16.21 & 20.75 & \boldy{25.51} & 12.51 & 16.95 & \underc{24.96} & 18.05 \\
Qwen2.5-7B-Instruct          & 17.29 & 24.88 & 19.71 & 18.24 & 15.37 & 17.91 & \boldy{25.31} & \underc{24.91} \\
Qwen2.5-Coder-7B-Instruct    & 20.39 & \boldy{33.00} & 22.60 & 31.21 & 17.47 & 21.39 & 30.29 & 27.74 \\
Qwen3-0.6B                   & 13.53 & 13.87 & 6.65  & \boldy{15.18} & 10.74 & 10.08 & \underc{14.26} & 12.77 \\
Qwen3-1.7B                   & 24.65 & 26.90 & 29.02 & \boldy{33.68} & 4.92  & 6.00  & \underc{30.76} & 29.03 \\
Qwen3-4B                     & 38.97 & 37.34 & \boldy{44.70} & 40.75 & 40.38 & 38.99 & 35.11 & \underc{43.13} \\
Qwen3-8B                     & 43.86 & 41.69 & 46.22 & 47.17 & 41.81 & 44.47 & 40.08 & \boldy{48.46} \\

\hline
\multicolumn{9}{c}{\textit{\textbf{Fine-tuned LLMs}}} \\
\hline
TableGPT2-7B                 & 14.11 & 22.78 & 15.65 & 15.25 & 17.88 & 20.34 & \boldy{26.65} & 20.45 \\
TableLLM-CodeQwen-7B         & 21.00 & \boldy{24.78} & 18.33 & 19.04 & 20.30 & 21.74 & 21.28 & 20.19 \\
TableLLM-DeepseekCoder-7B    & 22.59 & \boldy{30.06} & 21.65 & 27.21 & 24.78 & 21.81 & 21.77 & 25.00 \\
TableLLM-Llama3-8B           & 20.24 & \boldy{27.61} & 21.37 & 21.01 & 20.98 & 21.40 & 20.84 & 20.91 \\
TableLLM-Llama3.1-8B         & 22.22 & \boldy{28.40} & 21.00 & 25.68 & 21.09 & 22.72 & 21.29 & 21.43 \\
TableLLM-Qwen2-7B            & 22.72 & 31.81 & 24.40 & 30.78 & 30.52 & 27.02 & \boldy{34.41} & 30.20 \\

\hline
\end{tabular}
\vspace{-0.5cm}
\end{table}

\noindent After task-wise analysis, in Table~\ref{tab:tablebench_overall_result} we showcase the overall accuracy of methods on TableBench \cite{wu2025tablebench} dataset, where TGN demonstrated clear superiority, outperforming baselines including ReAct and CoT and achieving SOTA performance of 48.46\%. For tabular fine-tuned LLMs, PIP scored highest with an accuracy of 34.41\% and achieving highest in 5 and second highest score in 3 out of 17 LLMs.

\begin{table}[!ht]
\centering
\caption{Quantitative Analysis of metrics on FeTaQa test set by selection of best performance of each baseline. Here, R-1, R-2 and R-L in the table represents ROUGE-1, ROUGE-2 and ROUGE-L metrics respectively. Column best value is represented using Yellow (bold) and second best using Cyan (underline). }
\label{tab:fetaqa}
\tiny
\resizebox{\textwidth}{!}{%
\begin{tabular}{c | c c c c c c c}
\rowcolor{HeaderBlue}
\multicolumn{1}{c|}{\color{white}\textbf{Methodology}} &
\color{white}\textbf{sacreBLEU $\uparrow$} &
\color{white}\textbf{\hspace{0.4em}R-1 $\uparrow$\hspace{0.4em}} &
\color{white}\textbf{\hspace{0.4em}R-2 $\uparrow$\hspace{0.4em}} &
\color{white}\textbf{\hspace{0.4em}R-L $\uparrow$\hspace{0.4em}} &
\color{white}\textbf{METEOR $\uparrow$} &
\color{white}\textbf{\hspace{0.4em}BERTScore $\uparrow$\hspace{0.4em}} &
\color{white}\textbf{BLEURT $\uparrow$} \\

\hline
\addlinespace[0.2em]

Meta-Llama-3-8B-Instruct + DP & 
17.01 & 0.52 & 0.31 & 0.43 & 0.42 & \boldy{0.67} & 0.54 \\

DeepSeek-R1-Distill-Llama-8B + TCoT & 
13.91 & 0.47 & 0.27 & 0.38 & 0.38 & 0.44 & 0.51 \\

Llama-3.1-8B-Instruct + TCoT & 
14.83 & 0.47 & 0.28 & 0.39 & 0.38 & 0.42 & 0.51 \\

TableGPT2-7B + SCoT & 
15.37 & 0.48 & 0.29 & 0.39 & 0.40 & 0.44 & 0.58 \\

DeepSeek-R1-Distill-Llama-8B + ReAct & 
17.64 & 0.54 & 0.32 & 0.44 & 0.45 & 0.52 & 0.58 \\

Llama-3.1-8B-Instruct + ToT & 
17.28 & 0.54 & 0.32 & 0.44 & 0.44 & 0.51 & 0.58 \\

\hline

\textbf{Meta-Llama-3-8B-Instruct + PIP (Ours)} & 
\boldy{19.32} & \boldy{0.58} & \boldy{0.35} & \boldy{0.47} & \boldy{0.48} & \underc{0.56} & \boldy{0.60} \\

\textbf{Meta-Llama-3-8B-Instruct + TGN (Ours)} & 
\underc{17.96} & \underc{0.54} & \underc{0.33} & \underc{0.45} & \underc{0.45} & 0.52 & \underc{0.58} \\

\hline
\end{tabular}
}
\vspace{-0.5cm}
\end{table}

Next, we evaluate on FeTaQA \cite{nan2022fetaqa} test set, by running inference using 17 LLMs for each baseline, resulting a total of 136 inferences. To keep the results insightful and compact, we select the best performing output of each baseline across the LLMs, as shown in Table~\ref{tab:fetaqa}. We observe that the methodology using PIP framework achieved SOTA accuracy on FeTaQA \cite{nan2022fetaqa} with TGN as second best scorer and outperforming other baselines. In Table~\ref{tab:finding}, we further demonstrate the capability of our proposed framework on TQA task by comparing the overall accuracy of LLMs greater than 8B parameters with smaller LLM. Here, Qwen3-8B with TGN, having 48.46\% accuracy, surpasses many large models with baselines on the TableBench \cite{wu2025tablebench} dataset.

\begin{table}[!ht]
\centering
\caption{Overall accuracy comparison of language models with PIP and TGN framework against  models $\geq$ 8B parameters using baselines on TableBench dataset.}
\label{tab:finding}
\tiny
\setlength{\tabcolsep}{5pt}

\begin{tabular}{l | c c}
\rowcolor{HeaderBlue}
\multicolumn{1}{c|}{\color{white}\textbf{Methodology}} &
\color{white}\textbf{\# Params. $\downarrow$} &
\color{white}\textbf{Overall Acc. $\uparrow$} \\

\hline
GPT-3.5-Turbo + PoT & -    & 37.15 \\
Llama3-70B-Chat + TCoT & 70B & 38.68 \\
Llama3.1-70B-Instruct + TCoT & 70B & 41.05 \\
QWQ-32B + DP & 32B & 43.87 \\
Qwen2.5-Coder-32B-Instruct + TCoT & 32B & 45.51 \\
Llama-4-Scout-17B-16E-Instruct + TCoT & 17B & 46.53 \\
Qwen2.5-72B-Instruct + TCoT & 72B & 48.79 \\
Llama-3.1-405B-Instruct + TCoT & 405B & 48.87 \\
\textbf{DeepSeek-R1-Distill-Llama-8B + PIP (ours)} & \boldy{8B} & \boldy{37.48}\\
\textbf{Qwen3-8B + TGN (ours)} & \boldy{8B} & \boldy{48.46}\\

\hline
\end{tabular}
\vspace{-0.5cm}
\end{table}

\noindent These findings challenge the prevailing 'bigger is better' paradigm in LLM performance, showing that sophisticated framework enables compact models to transcend the accuracy benchmarks set by significantly larger models using conventional methods. For more detailed analysis, we share the accuracy of our proposed frameworks on different question categories present in TableBench dataset in Appendix~\ref{sec:appendix_detailed_analysis}.

\begin{table}[!ht]
\centering
\caption{Average token consumption of language models during inference with PIP and TGN framework against baselines on TableBench dataset.}
\label{tab:efficiency}
\tiny
\setlength{\tabcolsep}{5pt}

\begin{tabular}{c | c}
\rowcolor{HeaderBlue}
\multicolumn{1}{c|}{\color{white}\textbf{Framework}} &
\color{white}\textbf{Avg. Token Count / Example $\downarrow$} \\

\hline
SCoT & 1105.58 \\
ToT & 795.67 \\
ReAct & 788.09 \\
CoT & 778.83 \\
ToT\_SelfAsk & 713.23 \\
\textbf{PIP} & \boldy{794.09} \\
\textbf{TGN}	& \boldy{680.22} \\
\hline
\end{tabular}
\vspace{-0.5cm}
\end{table}

\noindent For cost-effectiveness analysis of TGN \& PIP we analyzed the inference-time token consumption across all 17 models with each framework as shown in Table~\ref{tab:efficiency}, where TGN required 680.22 tokens/query on average, outperforming competitive baselines such as ReAct (788.09 tokens) and SCoT (1105.58 tokens).

\subsection{Error Analysis}

To analyze systematic limitations, we conducted an analysis of failures across both datasets. In TableBench, while the model achieves high proficiency in Fact Checking and Numerical Reasoning it primarily fails in Data Analysis, where its subtypes require pattern interpretation over tables. Causal Analysis (38.13\%), Anomaly Detection (32.32\%), and Trend Forecasting (18.0\%) — all hover near chance, as the model produces plausible but table-unsupported reasoning; Statistical Analysis (28.0\%) and Descriptive Analysis (23.67\%) similarly fails at open-ended summarization, while  Impact Analysis (36.0\%) improve as expected outputs become more constrained. Within Numerical Reasoning, vulnerabilities also emerge in Domain-Specific tasks and Counting tasks due to missing background knowledge and multi-condition filtering omissions. In FeTaQA, model struggles most with questions that require multi-fact synthesis or inferential generation. Causal questions are the hardest, with approx. 30.8\% answered correctly, as the model produces surface-level responses unsupported by table evidence. Questions which requires enumeration fails because the model retrieves only the most salient match rather than aggregating all qualifying rows values into a complete answer. Comparative/Ranking questions suffer from both incorrect extremum identification and incomplete contextualisation of the result. Answer which includes person or entity shows bimodal behavior, either correct retrieval or a hallucinated name with no partial credit. Ultimately, analysis of the poorest-performing FeTaQA generations isolates partial or incomplete answers (18.9\% of samples) and partial hallucinations (12.6\% of samples) as the dominant error modalities, underscoring a persistent systemic limitation.

\subsection{Ablation Study}

To evaluate the effectiveness, we conducted the ablation studies by selectively removing key modules and by measuring the sensitivity of frameworks as shown in Table~\ref{tab:ablation}. In TGN, removing the last two modules caused the model to reason over tables with incorrect actions (Stage 1), while removing only the validation module led to hallucinations due to poor grounding (Stage 2), resulting in a significant performance drop from its SOTA results. Similarly, ablation results on PIP demonstrated the critical role of structural decomposition: removing schema identification and intermediate reasoning (Case 1) disrupted logical consistency, whereas removing row extraction and analysis (Case 2) reduced accuracy by preventing the model from isolating relevant variables from tabular noise.

\begin{table}
\centering
\caption{Ablation study of TGN \& PIP Framework respectively by performing inference on TableBench using top scoring LLMs from overall result.}
\label{tab:ablation}
\tiny
\setlength{\tabcolsep}{3pt}
\begin{tabular}{c | l | c c}
\hline
\rowcolor{HeaderBlue}
\multicolumn{2}{c|}{\color{white}\textbf{Model}} &
\color{white}\textbf{Stage 1} &
\color{white}\textbf{Stage 2} \\
\hline
\multirow{3}{*}{\textbf{TGN}}
& Qwen3-8B            & 41.6  & 47.17 \\
& Qwen3-4B            & 37.34 & 40.75 \\
& Qwen2.5-7B-Instruct & 24.88 & 18.24 \\
\hline

\rowcolor{HeaderBlue}
\multicolumn{2}{c|}{\color{white}\textbf{Model}} &
\color{white}\textbf{Case 1} &
\color{white}\textbf{Case 2} \\
\hline
\multirow{3}{*}{\textbf{PIP}}
& Qwen2.5-7B-Instruct          & 21.39 & 19.14 \\
& DeepSeek-R1-Distill-Llama-8B & 35.83 & 35.13 \\
& TableLLM-Qwen2-7B            & 31.79 & 33.29 \\
\hline
\end{tabular}
\vspace{-0.5cm}
\end{table}

\noindent For sensitivity, we evaluate TGN \& PIP using heavily paraphrased templates without disturbing the logical structure of the framework. Applying these altered templates to Qwen3-8B on TableBench showed remarkable stability, for TGN the accuracy shifted from 48.46\% to 45.45\% \& for PIP it remained identical from 40.08\% to 40.52\%. This minimal variance gives evidence that gains of TGN \& PIP are driven by logic of the framework.

\section{Conclusion}
\label{sec:conclusion}
In this paper, we introduced TGN and PIP, the first prompting framework designed to enhance the complex reasoning capabilities of LLMs in TableQA. By formalizing TGN as an iterative, validation-driven cycle for answer grounding from table and PIP as a constraint enforcer of column and row selection related to the query. We addressed key limitations in existing methods, such as unstructured reasoning and inefficiency in tabular data handling. Through comprehensive experiments on seventeen state-of-the-art large language models across two different categories, we demonstrate that TGN and PIP consistently outperformed baselines like Chain-of-Thought, and ReAct.

\begin{credits}
\subsubsection{\ackname}This research work was funded by Visvesvaraya PhD Scheme for Electronics \& IT Phase-II under the Ministry of Electronics and Information Technology, Government of India.

\subsubsection{\discintname}
The authors have no competing interests to declare that are relevant to the content of this article.
\end{credits}
%
%
%
\bibliographystyle{splncs04}
\bibliography{refs}

\clearpage
\appendix

\section{Prompts}
\label{sec:appendix_prompts}

In this section, we present demonstration used across TableBench dataset. We select the same answer output format from TableBench across multiple sub-tasks to keep the evaluation fair and the metrics on same scale as original.

\subsection{Demonstration of Progressive Inference Prompting}
As shown in Table~\ref{tab:pip_appendix}, our primary goal is to help the model understand the process of progressive inference workflow through the zero-shot five steps instructions given to it. The \{table\_str\} contains the full table string in JSON format after pre-processing and \{question\} contains the query for the given table, for which the LLM generates the answer in the given answer format.

\begin{table}[h!]
\caption{Default instruction template for Progressive Inference Prompting.}
\centering
\scriptsize
\begin{tabular}{ p{\linewidth} }
\hline
\textbf{Progressive Inference Prompting on TableBench Dataset} \\
\hline
You are a table reasoning assistant. You will be given a table and a question. Your goal is to progressively reason over the table before giving the final answer. \\ \\

[Guidelines] \\
You should act in following patterns step by step to analyze the table and then give the final answer: \\

[Action Patterns] \\
Step 1: Identify the table columns and their meaning. \\
Step 2: Identify the question and restate the question in your own words. \\
Step 3: Extract relevant rows from the table in a sequence. \\
Step 4: Do intermediate analysis, calculations, or comparisons step by step for each relevant row only once. \\
Step 5: Provide the final answer. \\ \\ \\

The answer should follow the format below: \\

[Answer Format] \\
Final Answer: AnswerName1, AnswerName2... \\ \\

Ensure the final answer format is the last output line and can only be in the "Final Answer: AnswerName1, AnswerName2..." form, no other form. Ensure the "AnswerName" is a number or entity name, as short as possible, without any explanation. \\ \\

[TABLE] \\
\{table\_str\} \\ \\

Let's get start! \\
Question: \{question\} \\

\hline
\end{tabular}
\label{tab:pip_appendix}
\end{table}

\subsection{Demonstration of TableGrid Navigation Prompting}
The prompting technique of TGN is shown in Table~\ref{tab:tgn_appendix} with three modules, where we tried to explain the model about how each module works and these modules can be repeated until the query is resolved. Here also the \{table\_str\} contains the full table string in JSON format after pre-processing and \{question\} contains the query for the given table.

\begin{table}[!ht]
\caption{Default instruction template for TableGrid Navigation.}
\centering
\scriptsize
\begin{tabular}{ p{\linewidth} }
\hline
\textbf{TableGrid Navigation Prompting on TableBench Dataset} \\
\hline
You are a table assistant who solves the query by analyzing the question and executing operations. \\ \\

[Guidelines] \\
Use the following format for processing tabular queries: \\ \\

Analyze: [reasoning about how to interpret the data grid or query] \\
Execute: [specific operation to perform on the tabular schema, e.g., lookup, calculation, or aggregation] \\
Validate: [verification of the result against the data grid] \\
... (repeat Analyze - Execute - Validate as needed to resolve the query) \\ \\ \\

The answer should follow the format below: \\

[Answer Format] \\
Final Answer: AnswerName1, AnswerName2... \\ \\

Ensure the final answer format is the last output line and can only be in the "Final Answer: AnswerName1, AnswerName2..." form, no other form. Ensure the "AnswerName" is a number or entity name, as short as possible, without any explanation. \\ \\

[TABLE] \\
\{table\_str\} \\ \\

Let's get start! \\
Question: \{question\} \\

\hline
\end{tabular}
\label{tab:tgn_appendix}
\end{table}

\subsection{Prompt used on FeTaQA Dataset}
Inference on the FeTaQA dataset was performed using the similar framework configurations specified in Tables \ref{tab:pip_appendix} and \ref{tab:tgn_appendix} for PIP and TGN, respectively. As FeTaQA employs free-form natural language answers rather than structured outputs in its ground truth annotations, corresponding adjustments were made to the inference configuration.

\begin{table}[h!]
\caption{Answer Format used on FeTaQa dataset for inference.}
\centering
\scriptsize
\begin{tabular}{ p{\linewidth} }
\hline
\textbf{Answer Format for FeTaQA Dataset} \\
\hline
The answer should follow the format below: \\

[Answer Format] \\
Final Answer: AnswerSentence \\ \\

Ensure the final answer format is the last output line and can only be in the "Final Answer: AnswerSentence" form, no other form. Ensure the "AnswerSentence" is a single sentence, as short as possible, without any explanation. \\ \\

[TABLE] \\
\{table\_str\} \\ \\

Let's get start! \\
Question: \{question\} \\

\hline
\end{tabular}
\label{tab:fetaqa_appendix}
\end{table}

\noindent Specifically, we set the [Answer Format] to AnswerSentence, as shown in Table~\ref{tab:fetaqa_appendix}, to ensure format compatibility between model outputs and reference answers, thereby enabling fair and meaningful evaluation.

\section{Detailed Analysis}
\label{sec:appendix_detailed_analysis}

In this section, we share the results of our proposed frameworks on different question categories present in TableBench \cite{wu2025tablebench} dataset. TGN accuracy on different question categories are shown in Table~\ref{tab:appendix_tgn_detailed1} \& \ref{tab:appendix_tgn_detailed2}, and for PIP we share the results in Table~\ref{tab:appendix_pip_detailed1} \& \ref{tab:appendix_pip_detailed2}. Abbreviations used in tables are shared below.

\begin{table}[!ht]
\caption{Quantitative analysis of accuracy across different question categories present in TableBench dataset using TGN framework (Part 1).}
\centering
\scriptsize
\setlength{\tabcolsep}{2.5pt}
\begin{tabular}{l|cccccccc}
\hline
\rowcolor{HeaderBlue}
\color{white}{Model} &
\color{white}{NR-A} &
\color{white}{FC-MB} &
\color{white}{DA-I} &
\color{white}{DA-C} &
\color{white}{NR-MH} &
\color{white}{FC-MH} &
\color{white}{NR-Cn} &
\color{white}{DA-Ds} \\
\hline
DeepSeek-R1-Distill-Llama-8B & 24 & 52.17 & 22 & 36.36 & 21.57 & 38 & 32 & 24.94 \\
Llama-3.1-8B-Instruct & 0 & 52.17 & 10 & 39.48 & 1.96 & 34 & 6 & 25.87 \\
Llama-3.2-3B-Instruct & 0 & 6.52 & 2 & 43.71 & 0 & 8 & 4 & 18.55 \\
Meta-Llama-3-8B-Instruct & 2 & 67.39 & 12 & 49.34 & 3.92 & 34 & 8 & 28.31 \\
Qwen2-7B-Instruct & 2 & 67.39 & 14 & 37.76 & 1.96 & 54 & 18 & 23.62 \\
Qwen2.5-7B-Instruct & 10 & 73.91 & 24 & 37.31 & 17.65 & 44 & 12 & 28.33 \\
Qwen2.5-Coder-7B-Instruct & 6 & 82.61 & 26 & 36.34 & 17.65 & 60 & 18 & 21.57 \\
Qwen3-0.6B & 18 & 26.09 & 4 & 35.77 & 7.84 & 10 & 10 & 14.78 \\
Qwen3-1.7B & 46 & 54.35 & 8 & 40.17 & 33.33 & 38 & 58 & 17.32 \\
Qwen3-4B & 48 & 97.83 & 18 & 30.28 & 41.18 & 68 & 72 & 25.34 \\
Qwen3-8B & 56 & 97.83 & 36 & 38.13 & 50.98 & 74 & 76 & 23.67 \\
TableGPT2-7B & 12 & 67.39 & 24 & 25.65 & 13.73 & 22 & 30 & 15.85 \\
TableLLM-CodeQwen-7B & 4 & 82.61 & 20 & 44.34 & 1.96 & 54 & 10 & 23.61 \\
TableLLM-DeepseekCoder-7B & 8 & 77.78 & 30.61 & 45.71 & 17.65 & 55.1 & 18.37 & 25.73 \\
TableLLM-Llama3-8B & 4 & 73.91 & 32 & 45.92 & 0 & 48 & 20 & 25.44 \\
TableLLM-Llama3.1-8B & 6 & 84.78 & 32 & 47.08 & 1.96 & 46 & 14 & 26.29 \\
TableLLM-Qwen2-7B & 12 & 73.91 & 32 & 45.4 & 19.61 & 50 & 42 & 26.06 \\
\hline
\end{tabular}
\label{tab:appendix_tgn_detailed1}
\end{table}

\begin{table}[!ht]
\caption{Quantitative analysis of accuracy across different question categories present in TableBench dataset using TGN framework (Part 2).}
\centering
\scriptsize
\setlength{\tabcolsep}{2.5pt}
\begin{tabular}{l|cccccccc}
\hline
\rowcolor{HeaderBlue}
\color{white}{Model} &
\color{white}{DA-An} &
\color{white}{NR-Dm} &
\color{white}{NR-T} &
\color{white}{NR-Ar} &
\color{white}{NR-Rk} &
\color{white}{DA-Tr} &
\color{white}{DA-St} &
\color{white}{NR-Cp} \\
\hline
DeepSeek-R1-Distill-Llama-8B & 23.43 & 38.78 & 38.3 & 30 & 34 & 20 & 6 & 30 \\
Llama-3.1-8B-Instruct & 23.62 & 16.33 & 12.77 & 0 & 24 & 22 & 0 & 20 \\
Llama-3.2-3B-Instruct & 15.39 & 12.24 & 4.26 & 0 & 10 & 10 & 0 & 0 \\
Meta-Llama-3-8B-Instruct & 25.49 & 16.33 & 14.89 & 12 & 22 & 26 & 2 & 18 \\
Qwen2-7B-Instruct & 16.75 & 8.16 & 10.64 & 12 & 22 & 18 & 4 & 20 \\
Qwen2.5-7B-Instruct & 28.36 & 20.41 & 23.4 & 32 & 34 & 16 & 12 & 40 \\
Qwen2.5-Coder-7B-Instruct & 22.57 & 32.65 & 27.66 & 30 & 40 & 16 & 16 & 52 \\
Qwen3-0.6B & 22.46 & 2.04 & 19.15 & 28 & 4 & 6 & 8 & 18 \\
Qwen3-1.7B & 24.14 & 22.45 & 38.3 & 46 & 24 & 18 & 16 & 38 \\
Qwen3-4B & 28.99 & 55.1 & 51.06 & 56 & 82 & 16 & 18 & 64 \\
Qwen3-8B & 32.32 & 55.1 & 65.96 & 74 & 74 & 18 & 28 & 64 \\
TableGPT2-7B & 12.23 & 22.45 & 17.02 & 28 & 22 & 18 & 6 & 36 \\
TableLLM-CodeQwen-7B & 33.86 & 12.24 & 12.77 & 6 & 26 & 18 & 8 & 10 \\
TableLLM-DeepseekCoder-7B & 38.67 & 20.41 & 19.15 & 10 & 34 & 16 & 10 & 28 \\
TableLLM-Llama3-8B & 35.14 & 18.37 & 12.77 & 6 & 22 & 14 & 6 & 20 \\
TableLLM-Llama3.1-8B & 36.73 & 16.33 & 10.64 & 6 & 22 & 20 & 6 & 18 \\
TableLLM-Qwen2-7B & 33.19 & 38.78 & 23.40 & 22 & 48 & 16 & 16 & 50 \\
\hline
\end{tabular}
\label{tab:appendix_tgn_detailed2}
\end{table}

\scriptsize
\noindent \textbf{Abbreviations:} 
NR = Numerical Reasoning; 
FC = Fact Checking; 
DA = Data Analysis; 
A = Aggregation;
MB = Match-Based;
I = Impact Analysis;
C = Causal Analysis;
MH = Multi-Hop;
Cn = Counting;
Ds = Descriptive Analysis;
An = Anomaly Detection;
Dm = Domain-Specific;
T = Time-based Calculation;
Ar = Arithmetic Calculation;
Rk = Ranking;
Tr = Trend Forecasting;
St = Statistical Analysis;
Cp = Comparison.

\begin{table}[!ht]
\caption{Quantitative analysis of accuracy across different question categories present in TableBench dataset using PIP framework (Part 1).}
\centering
\scriptsize
\setlength{\tabcolsep}{2.5pt}
\begin{tabular}{l|cccccccc}
\hline
\rowcolor{HeaderBlue}
\color{white}{Model} &
\color{white}{NR-A} &
\color{white}{FC-MB} &
\color{white}{DA-I} &
\color{white}{DA-C} &
\color{white}{NR-MH} &
\color{white}{FC-MH} &
\color{white}{NR-Cn} &
\color{white}{DA-Ds} \\
\hline
DeepSeek-R1-Distill-Llama-8B & 56 & 76.09 & 10 & 38.88 & 47.06 & 44 & 38 & 26.34 \\
Llama-3.1-8B-Instruct & 6 & 58.7 & 2 & 39.71 & 3.92 & 38 & 4 & 23.61 \\
Llama-3.2-3B-Instruct & 0 & 15.22 & 2 & 37.28 & 0 & 10 & 2 & 17.55 \\
Meta-Llama-3-8B-Instruct & 12 & 80.43 & 6 & 42.37 & 7.84 & 46 & 12 & 28.2 \\
Qwen2-7B-Instruct & 10 & 82.61 & 18 & 29.01 & 15.69 & 58 & 20 & 21.15 \\
Qwen2.5-7B-Instruct & 10 & 65.22 & 26 & 31.99 & 11.76 & 38 & 10 & 26.17 \\
Qwen2.5-Coder-7B-Instruct & 8 & 89.13 & 38 & 41.18 & 15.69 & 48 & 20 & 26.62 \\
Qwen3-0.6B & 20 & 34.78 & 2 & 35.42 & 5.88 & 22 & 16 & 5.26 \\
Qwen3-1.7B & 56 & 60.87 & 4 & 34.27 & 35.29 & 42 & 54 & 0.9 \\
Qwen3-4B & 36 & 93.48 & 18 & 31.61 & 17.65 & 60 & 32 & 26.3 \\
Qwen3-8B & 30 & 95.65 & 32 & 32.44 & 33.33 & 70 & 44 & 25.26 \\
TableGPT2-7B & 14 & 73.91 & 20 & 35.43 & 15.69 & 34 & 36 & 20.41 \\
TableLLM-CodeQwen-7B & 4 & 86.96 & 32 & 44.94 & 1.96 & 52 & 12 & 24.52 \\
TableLLM-DeepseekCoder-7B & 4 & 78.26 & 26 & 45.5 & 3.92 & 46 & 18 & 25.87 \\
TableLLM-Llama3-8B & 4 & 73.91 & 34 & 47.23 & 0 & 50 & 6 & 26.23 \\
TableLLM-Llama3.1-8B & 4 & 84.78 & 24 & 46.71 & 3.92 & 48 & 12 & 24.87 \\
TableLLM-Qwen2-7B & 14 & 84.78 & 30 & 46.69 & 21.57 & 44 & 52 & 27.78 \\
\hline
\end{tabular}
\label{tab:appendix_pip_detailed1}
\end{table}

\begin{table}[!ht]
\caption{Quantitative analysis of accuracy across different question categories present in TableBench dataset using PIP framework (Part 2)}
\centering
\scriptsize
\setlength{\tabcolsep}{2.5pt}
\begin{tabular}{l|cccccccc}
\hline
\rowcolor{HeaderBlue}
\color{white}{Model} &
\color{white}{DA-An} &
\color{white}{NR-Dm} &
\color{white}{NR-T} &
\color{white}{NR-Ar} &
\color{white}{NR-Rk} &
\color{white}{DA-Tr} &
\color{white}{DA-St} &
\color{white}{NR-Cp} \\
\hline
DeepSeek-R1-Distill-Llama-8B & 21.53 & 48.98 & 53.19 & 54 & 54 & 26 & 24 & 50 \\
Llama-3.1-8B-Instruct & 20.78 & 16.33 & 8.51 & 2 & 34 & 18 & 6 & 16 \\
Llama-3.2-3B-Instruct & 14.88 & 0 & 8.51 & 0 & 10 & 8 & 0 & 0 \\
Meta-Llama-3-8B-Instruct & 23.78 & 20.41 & 23.4 & 28 & 34 & 14 & 14 & 36 \\
Qwen2-7B-Instruct & 22.1 & 30.61 & 23.4 & 34 & 32 & 12 & 12 & 32 \\
Qwen2.5-7B-Instruct & 28.6 & 32.65 & 36.17 & 26 & 44 & 26 & 18 & 30 \\
Qwen2.5-Coder-7B-Instruct & 23.79 & 38.78 & 25.53 & 38 & 54 & 16 & 18 & 50 \\
Qwen3-0.6B & 20.61 & 4.08 & 19.15 & 28 & 8 & 10 & 8 & 22 \\
Qwen3-1.7B & 24.42 & 20.41 & 57.45 & 52 & 38 & 18 & 16 & 38 \\
Qwen3-4B & 34.55 & 55.1 & 36.17 & 44 & 72 & 14 & 14 & 44 \\
Qwen3-8B & 28.86 & 55.1 & 53.19 & 48 & 76 & 14 & 14 & 56 \\
TableGPT2-7B & 24.73 & 32.65 & 23.4 & 44 & 36 & 6 & 20 & 48 \\
TableLLM-CodeQwen-7B & 33.53 & 12.24 & 17.02 & 4 & 16 & 20 & 10 & 18 \\
TableLLM-DeepseekCoder-7B & 39.04 & 16.33 & 12.77 & 10 & 28 & 20 & 10 & 16 \\
TableLLM-Llama3-8B & 31.59 & 22.45 & 14.89 & 6 & 24 & 18 & 8 & 16 \\
TableLLM-Llama3.1-8B & 33.7 & 20.41 & 10.64 & 10 & 24 & 20 & 2 & 22 \\
TableLLM-Qwen2-7B & 33.49 & 38.78 & 38.3 & 30 & 60 & 30 & 20 & 54 \\
\hline
\end{tabular}
\label{tab:appendix_pip_detailed2}
\end{table}

\scriptsize
\noindent \textbf{Abbreviations:} 
NR = Numerical Reasoning; 
FC = Fact Checking; 
DA = Data Analysis; 
A = Aggregation;
MB = Match-Based;
I = Impact Analysis;
C = Causal Analysis;
MH = Multi-Hop;
Cn = Counting;
Ds = Descriptive Analysis;
An = Anomaly Detection;
Dm = Domain-Specific;
T = Time-based Calculation;
Ar = Arithmetic Calculation;
Rk = Ranking;
Tr = Trend Forecasting;
St = Statistical Analysis;
Cp = Comparison.

\end{document}